# Anisotropy induced spin re-orientation in chemically-modulated amorphous ferrimagnetic films

E. Kirk[1,2], C. Bull[3], S. Finizio[2], H. Sepehri-Amin[4], S. Wintz[2], A.K. Suszka[1,2], N.S. Bingham[1,2], P. Warnicke[2], K. Hono[4], P.W. Nutter[3], J. Raabe[2], G. Hrkac[5*], T. Thomson[3*], L.J. Heyderman[1,2]

[1]Laboratory for Mesoscopic Systems, Department of Materials, ETH Zurich, 8093 Zurich, Switzerland

[2]Paul Scherrer Institute, 5232 Villigen PSI, Switzerland

[3]Department of Computer Science, University of Manchester, Oxford Road, Manchester M13 9PL, UK

[4]Research Center for Magnetic and Spintronic Materials, National Institute for Materials Science, Tsukuba 305-0047, Japan

[5]College of Engineering, Mathematics and Physical Sciences, University of Exeter, North Park Road, Exeter, EX4 4QF, UK

Corresponding authors:

*thomas.thomson@manchester.ac.uk

*g.hrkac@exeter.ac.uk

## Abstract

The ability to tune the competition between the in-plane and out-of-plane orientation of magnetization provides a means to construct thermal sensors with a sharp spin re-orientation transition at specific temperatures. We have observed such a tuneable, temperature driven spin re-orientation in structurally amorphous, ferrimagnetic rare earth-transition metal (RE-TM) alloy thin films using scanning transmission X-ray microscopy (STXM) and magnetic measurements. The nature of the spin re-orientation transition in FeGd can be fully explained by a non-equilibrium, nanoscale modulation of the chemical composition of the films. This modulation leads to a magnetic domain pattern of nanoscale speckles superimposed on a background of in-plane domains that form Laudau configurations in micron-scale patterned elements. It is this speckle magnetic structure that gives rise to a sharp two step-reversal mechanism that is temperature dependent. The possibility to balance competing anisotropies through the temperature opens opportunities to create and manipulate topological spin textures.

# I. INTRODUCTION

Ferrimagnets, with two magnetically ordered sublattices that are coupled, offer opportunities to create a host of functional materials. A particularly fascinating class of these materials are amorphous rare earth - transition metal (RE-TM) alloys in thin film form with perpendicular magnetic anisotropy [1]. Recently, RE-TM alloy thin films have garnered significant attention as materials for ultra-fast, all-optical magnetic switching [2, 3, 4], as systems that can support topologically isolated structures such as Skyrmions [5, 6, 7] and as materials for spin-orbit torque devices [8]. In the past, their high perpendicular magnetic anisotropy and lack of grain boundaries have led them to be exploited for bubble memories [9] and used as magneto-optic recording media [10].

Particularly interesting is the ability to design RE-TM thin films where temperature can be used to control spin-reorientation [11, 12], saturation magnetization [13], magnetic anisotropy [14] and induce specific spin configurations [5]. Many RE-TM thin films have a strong perpendicular magnetic anisotropy, although the mechanism responsible for generating this anisotropy in nominally amorphous RE-TM thin films is still not fully explained with a number of competing hypotheses proposed for various RE-TM alloy thin films. Five possible mechanisms capable of inducing perpendicular anisotropy in amorphous RE-TM alloy thin films are typically considered: (i) dipolar interaction between atoms leading to pair ordering due to surface roughness [15], (ii) selective resputtering due to the different bonding strengths of RE and TM atoms leading to pair ordering [16] (iii) rare earth single-ion anisotropy [17], (iv) magneto-elastic coupling leading to bond-orientation anisotropy [18, 19] and (v) induced microstructural changes [20], with perhaps the most widely accepted being that of Harris et al. [21] who demonstrated anisotropic pair-pair correlations in TbFe using EXAFS similar to mechanism (ii). Here we explore a further hypothesis, previously proposed by Graves et al. [22], for a similar class of GdFeCo thin films, which is that of chemical phase separation. Indeed, the chemical phase separation is something that is frequently overlooked as the assumption is often made that amorphous films are homogeneous.

In this report, we choose FeGd as a prototype RE-TM amorphous, ferrimagnetic thin film system with a composition tunable compensation temperature [23, 24, 25, 26, 27]. We show that for an appropriately chosen composition, FeGd thin films support a spin reorientation transition using scanning transmission X-ray microscopy (STXM) and vibrating sample magnetometer (VSM) measurements, and demonstrate how the complex nanoscale composition of these films leads to a two-step hysteresis behaviour at temperatures beyond the spin-reorientation temperature. The intricate magnetic nature of these RE-TM alloys originates from a metastable thermodynamic equilibrium. In particular, complex spin dynamics and local changes in magnetic properties can be induced via temperature and/or fabrication processes. It should be mentioned that a spin reorientation on changing the temperature has previously been reported in ultra-thin Fe films [28, 29] and RE-TM alloy thin films [11, 30, 12]. However, the ability to control and tailor this transition, and the resulting macroscopic properties, including anisotropy and coercivity, sufficiently for use in devices such as sensors, requires a more detailed understanding of the relationship between the evolution of the spin orientation with temperature and the thickness-dependent nanoscale compositional structure. Here we provide a detailed explanation of the influence of the chemical composition on the existence and temperature dependence of a spin reorientation transition in these materials.

## II. EXPERIMENTAL

### A. Sample preparation

We prepared FeGd(t nm)/Ta (8 nm) films (where t is the FeGd thickness), by DC magnetron co-sputtering at room temperature from elemental Fe and Gd targets onto Si (100) substrates coated with a 95 nm LPCVD grown $Si_3N_4$ layer. The two thicknesses of film were sputtered under identical conditions resulting in an average composition of $Fe_{0.69}Gd_{0.31}$ at%. The composition was chosen to ensure that the spin reorientation occurred below room temperature thereby eliminating any possibility of recrystallization due to elevated measurement temperatures. The base pressure of the load-locked sputter chamber was better than $3.0 \times 10^{-9}$ mBar and the Ar sputtering gas pressure was $4.8 \times 10^{-3}$ mBar. Films with thicknesses of t = 20 nm and t = 40 nm were produced to represent the two key regimes in the evolution of the observed lateral chemical segregation. STXM measurements were undertaken on the 40 nm - thick films patterned into arrays of square-shaped microstructures magnetic elements with side-length 5 μm. For the patterning MMA/PMMA bilayer lift-off masks were spin-coated on 1 mm × 1 mm $Si_3N_4$ membranes, where the membrane thickness of 200 nm was chosen to allow the transmission of X-rays. After deposition of the films lift-off was performed in acetone. The separation of the square-shaped microstructures was sufficiently large to ensure that the influence of stray fields from neighbouring squares was negligible. Continuous film samples were used for the magnetometry measurements.

### B. Measurements

Magnetic measurements were performed using a Quantum Design MPMS3 SQUID magnetometer. The films were mounted on a quartz rod holder with the magnetic field applied in the sample plane. To characterise the spin reorientation, temperature dependent M(H) loops were measured with a maximum applied magnetic field of 4 kOe, which was sufficient to ensure that all switching events had taken place. For T = 180 K, as shown in supplementary information Fig. S1 [31], there is an abrupt reversal, which is associated with magnetization reversal along the in-plane (IP) easy axis. For temperatures above the spin reorientation transition, the easy axis becomes out of plane (OOP). Therefore, the M(H) measurements acquire the characteristics of a hard axis loop, with a slow increase in magnetization to saturation indicating a rotation of the magnetization, as shown by the T = 275 K data in supplementary information Fig. S1 [31]. At intermediate temperatures M(H) loops consist of a superposition of these two behaviours, and from these loops each component of magnetization can be estimated. Hence three magnetization values can be obtained, the total magnetization measured at 4 kOe, the in-plane magnetization and the out-of-plane magnetization. The difference in M(H) behaviour at different temperatures was confirmed by high-field M(H) measurements where the magnetic field was increased up to 70 kOe, as well as vector VSM measurements.

X-ray reflectivity (XRR) and perpendicular X-ray diffraction (XRD) measurements were undertaken using a Rigaku Smartlab X-Ray diffractometer, operating at the CuKα1 wavelength, λ = 1.540593(2) Å. XRR measurements were taken using a step size of 0.0004° over a 2-theta range of 0.1–7.0° at rate of 0.01°/minute. For XRD measurements, a Ge(220) double bounce monochromator was used to measure diffraction spectra with a step size of 0.02°, over a 2-theta range of 20.0-80.0° at a rate of 0.1°/minute. Measured reflectivity data was fitted to a simulated curve generated from a defined structural model of the thin film using the Parrett recursive formalism [32] implemented in the GenX reflectivity package [33] in order to obtain the depth-dependent sample structure. The Scattering Length Density (SLD) profile from this model was parametrized to obtain the film thickness, interfacial root mean square (RMS) roughness and density (SLD) values for individual layers in each sample.

Scanning transmission electron microscopy (STEM) was performed using a Titan G2 80-200 microscope with a probe aberration corrector. The specimens for the STEM analysis were prepared using an FEI Helios G4-UX dual-beam system with the lift out method. The distribution of constituent elements and chemical composition of the films were measured using energy-dispersive X-ray spectroscopy (EDS).

Scanning Transmission X-ray Microscopy (STXM) measurements were undertaken at the PolLux beamline (X07DA), Swiss Light Source, Paul Scherrer Institute [34] on membrane samples. Magnetic contrast was obtained using X-ray magnetic circular dichroism (XMCD) that results from a differential absorption of circularly polarized X-rays for magnetization parallel or antiparallel to the beam propagation direction. The specimens were imaged both in the normal incidence geometry, sensitive to OOP magnetization, and with the sample at an angle of 30° to the beam. For an angle of 30°, the components of magnetization in the domains with both the OOP and IP magnetization could be imaged. Images were taken with the X-ray energy tuned to the Gd $M_5$ absorption edge at 1190 eV. A series of images of a 5 µm square-shaped microstructured FeGd element were recorded at temperatures between 150 K and 240 K, which spans the temperature range over which the spin reorientation transition took place.

## III. RESULTS AND DISCUSSION

The temperature dependence of the spin orientation in the two FeGd RE-TM thin films is shown in Fig. 1(a) and (b). The data were obtained from in-plane M(H) loops measured by SQUID magnetometry over the temperature range 150 - 300 K with a maximum applied field of 4 kOe as described in the Measurements section II.B. This data shows that manipulation of the spin reorientation temperature is possible in FeGd thin films. In particular, it can be seen that the spin reorientation starts at 250 K for the 20 nm film and shifts towards a lower temperature of 190 K for the 40 nm film. The temperature dependence of magnetization for both samples, measured using an applied field of 70 kOe, is shown in the supplementary information, Fig. S2 [31]. These data demonstrate a similar behavior in saturation magnetization (Ms) vs temperature (T) for both samples. The value of Ms shows some variation between samples which, given the identical sputtering conditions employed, is unexpected and most likely reflects the sensitivity of the system to chemical inhomogeneities. In order to clarify the nature of the spin reorientation, XMCD-STXM images of the magnetic configuration of FeGd elements with 40 nm thickness were recorded. The STXM images presented in Fig. 1(c) were taken with the sample normal oriented at 30° to the X-ray beam so that both in-plane and out-of-plane components of the magnetization can be observed. The pixel size of the STXM images is ≈50 nm providing nanoscale observations of the evolution of the domain structure during the spin re-orientation transition. Specifically, at 164 K, micrometer-sized in-plane domains with vortices are present in the films. The persistent OOP frame around the membranes was observed for all FeGd samples measured and is likely to be related to Gd enrichment, as indicated by XAS measurements, under the deep lithographic lift-off edge. This boundary region did not affect the STXM data obtained from the membrane. In addition, all other measurements were undertaken on continuous films and are therefore unaffected by this boundary region. As the temperature is increased to 177 K, a fine domain structure consisting of small domains starts to emerge. We estimate the diameter of these domains to be between 100 and 150 nm but note there is a significant uncertainty in this determination, see supplementary information Fig. S3 and discussion [31]. At 186 K, fine scale OOP domains are observed to nucleate homogeneously within the IP domains. The IP structure of micrometer-sized domains and vortices continues to co-exist alongside the fine scale OOP domain structures at 199 K, with the direction of magnetization in the new OOP domains is observed to be strongly correlated to the direction of magnetization in the previous IP domains. In particular, the greyscale contrast associated with circulation of the IP magnetization around vortex cores is completely replaced by the black/white contrast of OOP domains at 211 K. Further increasing the temperature leads to large connected OOP domains above 240 K. This interpretation of the temperature evolution of the domain structure was confirmed by additional measurements taken in the purely out-of-plane geometry, supplementary information, Fig. S4 [31].

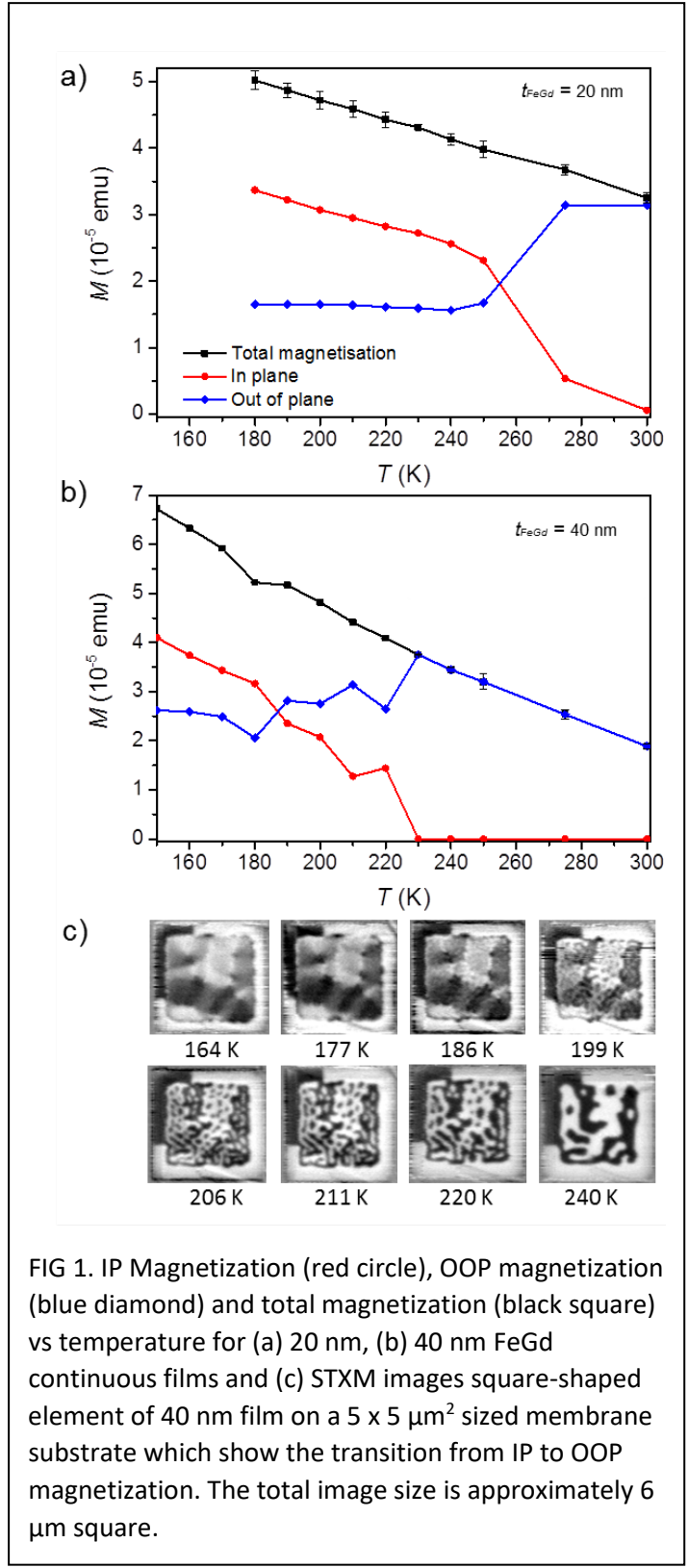

FIG 1. IP Magnetization (red circle), OOP magnetization (blue diamond) and total magnetization (black square) vs temperature for (a) 20 nm, (b) 40 nm FeGd continuous films and (c) STXM images square-shaped element of 40 nm film on a 5 x 5 µm$^2$ sized membrane substrate which show the transition from IP to OOP magnetization. The total image size is approximately 6 µm square.

The observed spin re-orientation can be understood in terms of a competition between the magnetic energy contributions. In the case of homogeneous films, the transition from the IP to OOP spin configuration is a function of the competition between demagnetization energy (volume and shape anisotropy), magnetic anisotropy energy and exchange energy. In a magnetic hybrid structure, where one encounters two or more different regions with different magnetic properties, more complex spin configurations can occur. The underlying phenomenon responsible for the complex magnetic configurations, as seen in Fig. 1(c), is competing energies on a local level, such as localized changes in magnetic anisotropy or variations in the magnetization resulting from an inhomogeneous chemical composition.

To fully understand the details of the observed magnetic behaviour we adopted an approach combining microstructural analysis and micromagnetic simulation. First, we undertook comprehensive measurements of the microstructural properties of the films using X-Ray diffraction (XRD) and Transmission Electron Microscopy (TEM). These data, shown in supplementary information Fig. S5 [31], demonstrate that the films are structurally amorphous, both locally as determined by TEM, and over larger length scales as shown by XRD. Therefore, a mechanism such as the pseudo-crystalline short-range order, proposed by Onton et al. [35] to describe the broadening of low energy XRD peaks, is not responsible for the perpendicular magnetic anisotropy observed in our films. We find instead that the underlying phenomena is that of chemical segregation as revealed by energy dispersive X-ray spectroscopy scanning transmission electron microscopy (STEM-EDS) and X-ray reflectivity (XRR) measurements (see methods). Chemical mapping of cross-sectional STEM-EDS for both the 20 nm and 40 nm samples is shown in Fig. 2(a), (b), (e) and (f), and the measured and simulated XRR data with the corresponding scattering length density (SLD) data are shown in Fig. 2(c), (d) ,(g) and (h). The experimental datasets could be successfully simulated using four and five layer models as shown in Fig. 2(c), (d) and (g), (h), respectively, with numerical values obtained from the fits given in the supplementary information, table SI [31]. Inclusion in the model of an oxidised Ta layer at the surface was essential to accurately reproduce the critical edge scattering in all samples investigated. The XRR Scattering Length Density (SLD) data clearly demonstrate a change in density through the thickness of the film, as demonstrated in Fig. 2(d) and Fig. 2(h). The STEM-EDS data provide further insight; the STEM results do not provide any evidence of induced voids [36, 37], which could potentially account for the density variation determined from the XRR analysis. The STEM-EDS maps show a compositional segregation occurs in the 40 nm film which is absent in the 20 nm film. In the case of the 40 nm film, in the first 20 nm above the substrate, there is a uniform composition distribution with 68.8 at% Fe and 31.2 at% Gd, which is close to the nominal composition of the films [Fig. 2(f(i))]. The standard deviation in the composition is 1.4 at%. However, a periodic modulation in chemical composition emerges as the film thickness increases as is evident from Fig. 2(fii), which is reflected by the fact that the standard deviation increases to 4.8 at%. This segregation is accompanied by a small change in the average composition, which increases from Gd = 31.2 at% to Gd = 35.4 at % with a commensurate reduction in Fe concentration, which is possibly due to the diffusion of Fe into Ta at the interfacial region. The chemical segregation leads to phase separation of ferrimagnetic properties, which is due to increased antiferromagnetic pair bonds between Gd-Fe relative to Fe-Fe pair bonds. This compositional change with thickness provides a convincing explanation for the change in scattering length density determined from XRR measurements. In addition, the chemically induced columnar structure, seen in the periodic change of the cross-sectional STEM images, leads to a magnetic easy axis parallel to the columns enriched with Gd embedded in an Fe-rich matrix due to the induced "shape anisotropy". Together, these data clearly show a chemical phase separation. Interestingly, this induced phase-separation and the associated preference for perpendicular magnetization only becomes established above a film thickness of 20 nm. Given the time to deposit 20 nm of FeGd (10 mins) it is possible to speculate that this is a result of the gradual temperature increase that occurs naturally during sputtering.

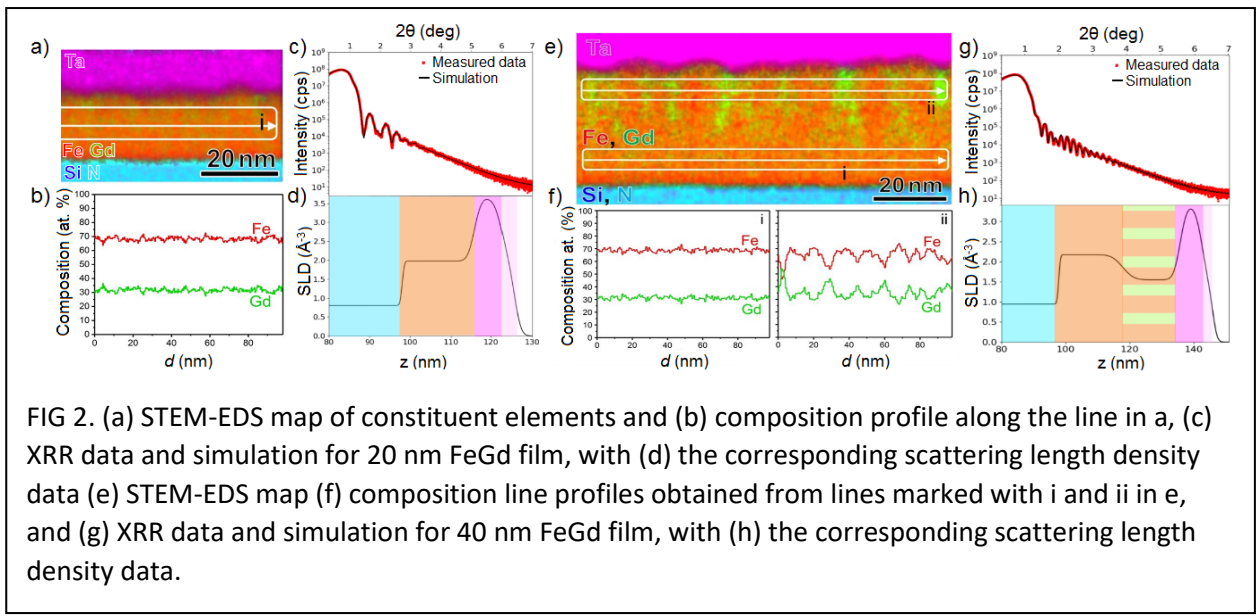

FIG 2. (a) STEM-EDS map of constituent elements and (b) composition profile along the line in a, (c) XRR data and simulation for 20 nm FeGd film, with (d) the corresponding scattering length density data (e) STEM-EDS map (f) composition line profiles obtained from lines marked with i and ii in e, and (g) XRR data and simulation for 40 nm FeGd film, with (h) the corresponding scattering length density data.

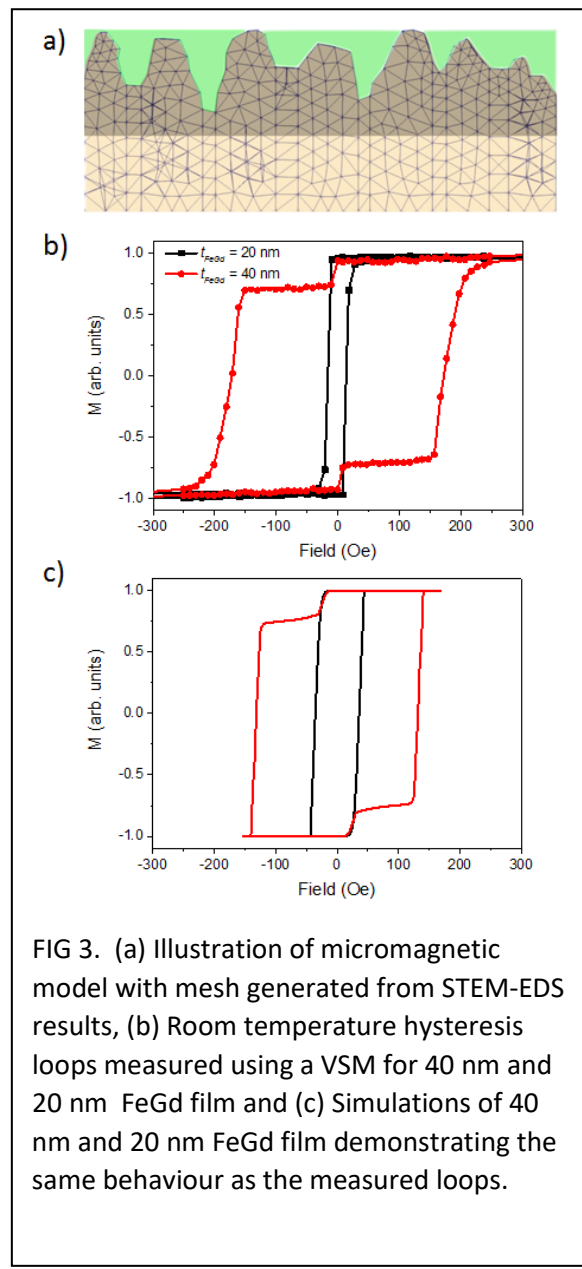

FIG 3. (a) Illustration of micromagnetic model with mesh generated from STEM-EDS results, (b) Room temperature hysteresis loops measured using a VSM for 40 nm and 20 nm FeGd film and (c) Simulations of 40 nm and 20 nm FeGd film demonstrating the same behaviour as the measured loops.

As can be seen from the SQUID and STXM measurements in Fig. 1, the chemical separation has an effect on the spin reorientation temperature with a clear reduction for the 40 nm film. To clarify the quantitative and qualitative effects of this chemical segregation, we measured out-of-plane hysteresis loops of the 20 nm and 40 nm FeGd films at 295 K (see Fig. 3(b)), which show that the magnetization is now perpendicular to the film plane. As expected, for the 20 nm film there is a single-step reversal behaviour with a coercivity of 15 Oe which is associated with the perpendicular anisotropy due to anisotropic pair-pair correlations. For the 40 nm film, the hysteresis loop is more complex with a two-step reversal. Inferring the behaviour from the STEM-EDS and STXM measurements, we associate the second reversal step with the switching of the columnar structure, which gives rise to the formation of the speckled domain structure seen in Fig. 1(c).

To test the hypothesis that the point domain structure is responsible for the two-step reversal, we performed micromagnetic simulations, implementing a three-dimensional finite element / boundary element micromagnetic columnar model based on the measured the composition line profiles as shown in Fig. 3(a) which allowed construction of a multi-phase FEM model. The effective structure in the 40 nm film consists of a 20 nm-thick layer of $Fe_2Gd$ and, on top of that, a 20 nm complex columnar structure. Although we constructed a columnar model specifically from the line profile, it should be noted that a simplified cylindrical column model with equal spacing also reproduces the two-step reversal behaviour. The simulations reproduce qualitatively the two-step reversal seen in the experiments. The difference in coercivity values is explained by the difference in the size of the simulated model and the experimental sample. The material parameters used in the micromagnetic simulations are $Ms(Fe_2Gd) = 0.369$ T, $Ms(FeGd) = 0.168$ T, $A = 1.e^{-12}$ J/m and are based on the calculation of the total Gibbs free energy [38].

Simulations of the hysteresis loops at room temperature reproduce the two-step reversal seen in the experimental data in Fig. 3(b). Indeed, without this chemically segregated columnar structure, the simulations fail to reproduce the two-step reversal observed experimentally. Above the spin reorientation temperature, two distinct hysteretic behaviours are observed; a single step loop with a coercivity of 15 Oe for the 20 nm and a two-step loop with coercivities of 5 Oe and 170 Oe for the 40 nm film.

## IV. CONCLUSIONS

We have deduced the origin of the speckled domain magnetization pattern observed in the STXM images. In particular, we have shown using STEM-EDS and XRR that chemical segregation leads to the emergence of a columnar anisotropy in Gd rich regions, which gives rise to the speckled domain structure and this is the driving force for the domain evolution in the 40 nm-thick film. This secondary anisotropy phase, associated with the Gd-rich columns, lowers the spin reorientation temperature. The presence of the columns also narrows temperature range over which the spin reorientation transition occurs, demonstrating the possibility to control spin reorientation transition temperature by manipulating the composition and microstructure. In this way, we achieve a higher functionality ferrimagnetic material that not only can be used as soft ferrimagnetic reference sensor that exploits the first low field transition, but also acts as temperature sensor. Additionally, the ability to tailor the balance between IP and OOP anisotropies allows specific spin textures to form, opening exciting new possibilities for creating novel temperature-dependent Skyrmionic systems.

Acknowledgements

The authors wish to acknowledge Vitaliy Guzenko and Anja Weber at PSI, Switzerland for the electron beam lithography to pattern the lift-off mask for the arrays of microstructured squares on the membranes. We acknowledge the Paul Scherrer Institut, Villigen, Switzerland, for provision of synchrotron radiation beamtime at X07DA (PolLux). The authors also gratefully acknowledge the contribution of the facilities of the Henry Royce Institute through EPSRC grants EP/S019367/1 and EP/P025021/1 for the X-ray reflectivity measurements.

Author contributions

EK and LH conceived the initial idea for project, with EK, TT and GH taking overall responsibility for the project work. EK conceived the experiment, sputter-deposited the FeGd films and performed the lithography to create the FeGd arrays. EK measured temperature-dependent M(H) loops and led the STXM and XMCD investigation of the temperature-dependent changes in the magnetic domain structure of FeGd film square-shaped microstructures.

JR and SF were responsible for the PolLux beamline (X07DA) where the STXM measurements were made and contributed to the data processing and analysis. SW, NB, AS and PW also contributed to the STXM measurements.

CB undertook the XRR measurements and data analysis. HS-A and KH performed the TEM measurements and analysis. GH performed all the micromagnetic simulations.

EK, CB, GH and TT were responsible for developing the analysis, explaining the results and coordinating the work. They were also responsible for drafting the paper to which all authors subsequently contributed. EK, LH, GH and TT completed the final form of the paper.